\documentclass[10pt, preprint]{aastex}
\usepackage{natbib}

\newcommand{\Wright}{{E. L. Wright}}

\newcommand{\UCLA}{{UCLA Astronomy, PO Box 951562, Los Angeles, CA 90095-1562}}

\singlespace

\shortauthors{Wright}

\shorttitle{WMAP First Year Results}

\newcommand{\iMAP}         {{\sl WMAP}}

\slugcomment{Draft: 30 May 2003}

\begin{document}

\title{The WMAP\altaffilmark{1}\ Data and Results}

\author{
\Wright \altaffilmark{2},
}

\altaffiltext{1}{\iMAP\ is the result of a partnership between Princeton 
                 University and NASA's Goddard Space Flight Center. Scientific 
		 guidance is provided by the \iMAP\ Science Team.}
\altaffiltext{2}{\UCLA}

\email{wright@astro.ucla.edu}

\begin{abstract}
The Wilkinson Microwave Anisotropy Probe (\iMAP) science team
has released results
from the first year of operation at the Earth-Sun L$_2$ Lagrange point.
The maps are consistent with previous observations but have much better
sensitivity and angular resolution than the {\sl COBE} DMR maps, and much
better calibration accuracy and sky coverage than ground-based and 
balloon-borne experiments.  The angular power spectra from these
ground-based and balloon-borne
experiments are consistent within their systematic and statistical
uncertainties with the \iMAP\ results.
\iMAP\ detected the large angular-scale correlation between the
temperature and polarization anisotropies of the CMB caused by
electron scattering since the Universe became reionized after the
``Dark Ages'', giving a value for the electron scattering optical depth
of $0.17 \pm 0.04$.
The simplest $\Lambda$CDM model with $n=1$ and $\Omega_{tot}=1$ fixed
provides an adequate fit to the \iMAP\ data and gives parameters which
are consistent with determinations of the Hubble constant and observations
of the accelerating Universe using supernovae.
The time-ordered data, maps, and power spectra from \iMAP\ can be found
at \verb"http://lambda.gsfc.nasa.gov" along with 13 papers by the
\iMAP\ science team describing the results in detail.
\end{abstract}


\newpage

\section{INTRODUCTION}\label{intro}

\begin{figure}[t]
\framebox[\textwidth]{\rule[-0.25\textwidth]{0in}{0.5\textwidth}\hfil
see 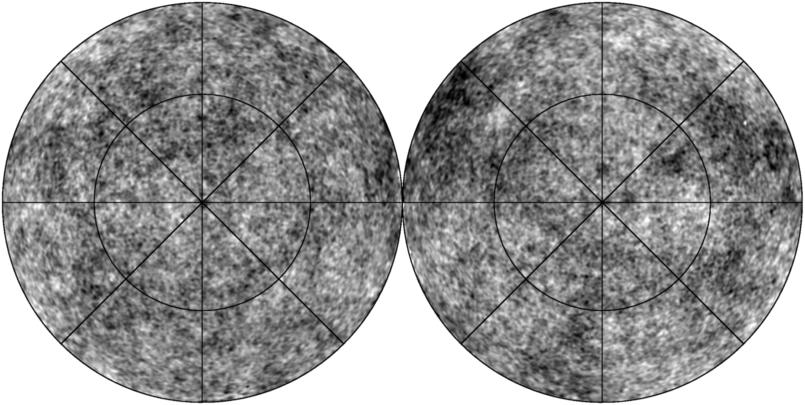 \hfil}
\caption{Internal linear combination reduced galaxy map from the \iMAP\ data.
This map is smoothed to about $1^\circ$ FWHM.\label{fig:ilc}}
\end{figure}

\begin{figure}[t]
\framebox[\textwidth]{\rule[-0.25\textwidth]{0in}{0.5\textwidth}\hfil
see 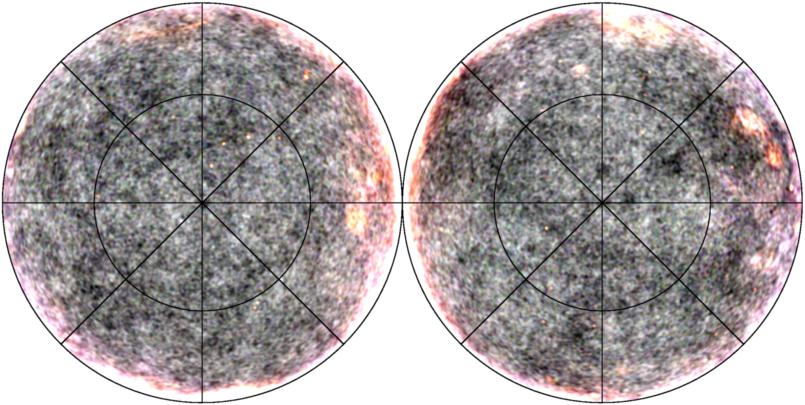 \hfil}
\caption{A red-green-blue color map made from the \iMAP\ Q, V \& W band
data.
This map is smoothed to about $1^\circ$ FWHM.\label{fig:QVW}}
\end{figure}

\begin{figure}[t]
\plotone{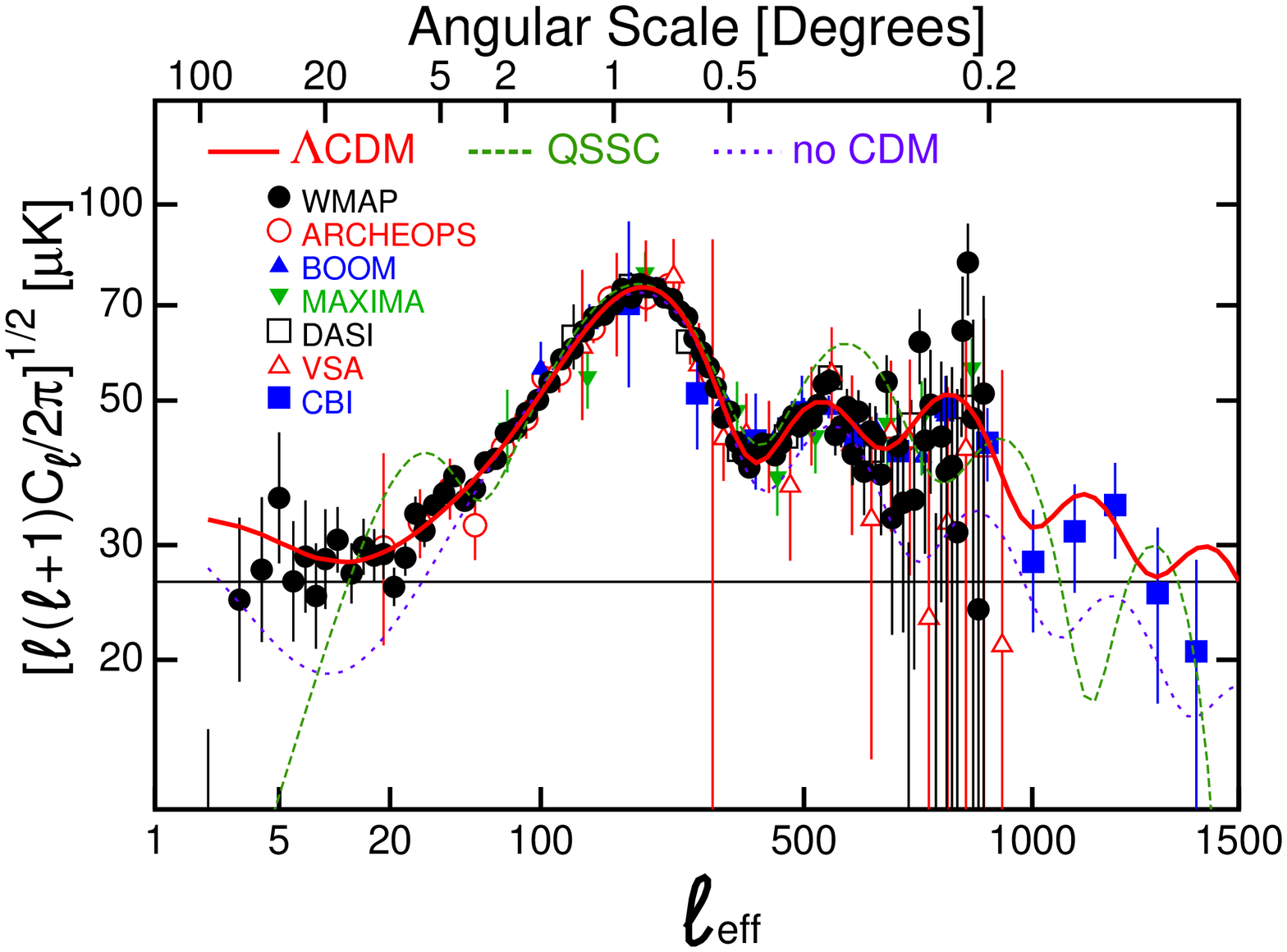}
\caption{The angular power spectrum of the CMB from \iMAP\ and earlier
balloon-borne and ground-based data.  The $\Lambda$CDM model is a good
fit to all of these datasets.
The Quasi-Steady-State cosmology and the no-CDM model inspired by MOND
(Modification Of Newtonian Dynamics) both give unacceptable best fits to
the CMB angular power spectrum, with large deviations in the low $\ell$
region observed by {\sl COBE} and confirmed by \iMAP.\label{fig:fits}}
\end{figure}

The cosmic microwave background (CMB) radiation was discovered by
\citet{penzias/wilson:1965}.  After its discovery, a small number of
experimentalists worked for years to better characterize the spectrum
of the CMB and to search for anisotropy in the CMB temperature.  A
leader of this effort, and of the \iMAP\ effort, was our recently
deceased colleague, Professor David T. Wilkinson of Princeton
University.  He was the supervisor of the doctoral theses that led to
the second \citep{henry:1971} and third \citep{corey/wilkinson:1976}
measurements of the dipole anisotropy of the CMB caused by the Solar
System's motion relative to the Universe.  He was also a leading
member of the {\sl Cosmic Background Explorer (COBE)} mission team,
which accurately characterized the spectrum of the CMB
\citep{mather/etal:1990, mather/etal:1999} and first discovered the
intrinsic (non-dipole) anisotropy \citep{smoot/etal:1992, bennett/etal:1992b,
kogut/etal:1992, wright/etal:1992} of the CMB.
The \iMAP\ science working group, led by Principal Investigator Charles L.
Bennett of Goddard Space Flight Center, was happy to have the opportunity
to honor David T. Wilkinson by renaming the Microwave Anisotropy Probe as
the Wilkinson Microwave Anisotropy Probe.

\citet{bennett/etal:2003} gives a description of the \iMAP\ mission.
\citet{bennett/etal:2003b} summarizes the results from first year of
\iMAP\ observations.
\citet{bennett/etal:2003c} describes the observations of galactic and
extragalactic foreground sources.
\citet{hinshaw/etal:2003} gives the angular power spectrum derived from the
the \iMAP\ maps.
\citet{hinshaw/etal:2003b} describes the \iMAP\ data processing and systematic
error limits.
\citet{page/etal:2003b} discusses the beam sizes and window functions
for the \iMAP\ experiment.
\citet{page/etal:2003c} discusses results that can be derived simply from
the positions and heights of the peaks and valleys in the angular power
spectrum.
\citet{spergel/etal:2003} describes the cosmological parameters derived
by fitting the \iMAP\ data and other datasets.
\citet{verde/etal:2003} describes the fitting methods used.
\citet{peiris/etal:2003} describes the consequences of the \iMAP\ results
for inflationary models.
\citet{jarosik/etal:2003b} describes the on-orbit performance of the \iMAP\
radiometers.
\citet{kogut/etal:2003} describes the \iMAP\ observations of polarization
in the CMB.
\citet{barnes/etal:2003} describes the large angle sidelobes of the
\iMAP\ telescopes.
\citet{komatsu/etal:2003} addresses the limits on non-Gaussianity that can be
derived from the \iMAP\ data.

\section{OBSERVATIONS \label{sec:obs}}

\iMAP\ Observatory was launched on 30 June 2001 at 19:46:46.183 UTC from
Cape Canaveral by a Delta expendable launch vehicle.  
After three phasing loops in the Earth-Moon system
\iMAP\ executed a lunar-gravity-assist swingby one month after launch
which catapulted \iMAP\ to an orbit about the second Lagrange point of
the Sun-Earth system, L$_2$.  L$_2$ is only metastable so about 4
station-keeping maneuvers per year are required to keep \iMAP\ on
station.  The spacecraft is actually in a ``halo'' orbit around L$_2$
and thus avoids the deep partial eclipse of the Sun that exists at
L$_2$.

\iMAP\ observes at 5 different frequencies and can thus yield internal
linear combination ``no galaxy maps''.  Figure \ref{fig:ilc} shows such
a map in the ``polar eyeball'' projection.  This equal-area projection
maps the North and South galactic hemispheres into circles with the
Galactic center in the middle, the North Galactic Pole (NGP) in the
center of the left circle, and the SGP in the center of the right
circle.  Figure \ref{fig:QVW} shows the 41, 61 \& 94 GHz maps as red,
green \& blue on the same temperature scale in the same projection as
Figure \ref{fig:ilc}.

\section{Analysis\label{sec:analysis}}

Given the extensive analysis of the \iMAP\ data already posted on the
astro-ph preprint server, or at the Legacy Archive for Microwave
Background Data Analysis (LAMBDA at http://lambda.gsfc.nasa.gov), I see
little point in repeating this lengthy discussion here.  I have
instead included Figure \ref{fig:fits} which shows a $\Lambda$CDM model
with a power law primordial power spectrum and zero spatial curvature.
This particular model was the best fit found in a fairly small Monte
Carlo Markov Chain computation that I did as an independent check of
the main \iMAP\ analysis papers.  One easily sees that the \iMAP\ data
are quite consistent with the previous experiments 
[ARCHEOPS \citep{benoit/etal:2003}, 
BOOMERanG \citep{debernardis/etal:2000},
DASI \citep{halverson/etal:2002}, MAXIMA \citep{hanany/etal:2000},
VSA \citep{grainge/etal:2003} \&
CBI \citep{padin/etal:2001}], some of which overlap with
\iMAP\ in $\ell$-space, and that the $\Lambda$CDM model is consistent
with both the \iMAP\ data and the higher angular resolution
interferometric data from CBI.

The largest discrepancies among various CMB anisotropy experiments are
due to systematic calibration uncertainties.  Therefore, in doing these
fits, a calibration correction for each experiment other than WMAP has
been introduced as a new parameter in the fits, and the sum of the
squares of calibration corrections divided by their stated
uncertainties has been added to the $\chi^2$ of the model fit.  None of
the calibration corrections is inconsistent with its stated
uncertainty.  The data are plotted with the calibration corrections
applied, which emphasizes the concordance between recent measurments of
$C_\ell$.

This plot is a good way to verify that \iMAP\ is consistent with
earlier experiments, but when setting limits on cosmological parameters
one should not combine \iMAP\ with experiments like ARCHEOPS that cover
the same $\ell$ range since the cosmic variance is correlated between
the two datasets.  Combining \iMAP\ only with high angular resolution
datasets like ACBAR \citep{kuo/etal:2002} and CBI is the correct
procedure.


I then used the MCMC code to optimize the parameters for two
alternative cosmological models that have published claimed fits to the
CMB angular power spectrum.  The ``no CDM'' model is an update of the
\citet{mcgaugh:2000} fit to the BOOMERanG data using CMBfast 
\citep{seljak/zaldarriaga:1996} with a
zero CDM density.  It is clear that this model is a terrible fit, and
it was a terrible fit to the combined CMB dataset including COBE which
existed in 2000.  If CMBfast with no CDM were a good predictor of the
anisotropy expected from the Modification Of Newtonian Dynamics then
MOND would be killed by this bad fit.  But CMBfast assumes that general
relativity is a good description of gravity and thus
\citet{mcgaugh:2000} is not a definitive prediction of the $C_\ell$
expected under MOND.
Thus the failure of the \citet{mcgaugh:2000} model to fit the data
only rules out this specific attempt to extend MOND to the early Universe.

More recently
\citet{narlikar/etal:2003} claim to have calculated the CMB anisotropy
expected in the Quasi-Steady-State Cosmology (QSSC), and also claim
that the QSSC gives a better fit to the {\em binned} data than
$\Lambda$CDM.  
This model is an {\it ad hoc} superposition of two populations of
Gaussian blobs with two different sizes and a population of hard-edged
circular spots all of a single size.
These hard-edged circular disks have an oscillatory Fourier transform
which gives a series of peaks in the angular power spectrum.
But the low-$\ell$ behavior of this model is $C_\ell =$~const, which
corresponds to a primordial power spectrum $P(k) \propto k^n$ with a power
law index of $n=3$ which is $6\sigma$
away from the COBE value of $n = 1.2 \pm 0.3$ \citep{bennett/etal:1996}.
\citet{narlikar/etal:2003} hide this failing of their model by using a
binning which puts all the COBE data into one bin.
I have re-optimized the
six arbitrary parameters in the \citet{narlikar/etal:2003} model to
give the best possible fit to the data in Figure \ref{fig:fits}, but
even the best fit is totally unacceptable.  As with MOND, if
\citet{narlikar/etal:2003} had presented a valid theory for the anisotropy
predicted by the QSSC, then this bad fit would have killed the QSSC.
But unfortunately neither the theory nor the fit is acceptable.

\section{Summary and Conclusions}

\iMAP\ has presented the results from its first year of observation at
L$_2$, and these data agree with the concordance $\Lambda$CDM model and
yield dramatic improvements in the accuracy of the cosmological
parameters.  \iMAP\ is funded for 3 more years of operation, and the 
results from a 4 year dataset will have a much improved signal-to-noise
ratio for $\ell > 400$ in the temperature anisotropy angular power spectrum,
and a much improved SNR for all $\ell$'s in the polarization measurements.

\acknowledgements

The \iMAP\ mission is made possible by the support of the Office of Space 
Sciences at NASA Headquarters and by the hard and capable work of scores of 
scientists, engineers, technicians, machinists, data analysts, budget analysts, 
managers, administrative staff, and reviewers. 

\bibliographystyle{apj}

\end{document}